\documentclass[fleqn,10pt]{wlscirep}
\usepackage[utf8]{inputenc}
\usepackage[T1]{fontenc}
\title{Water and heat exchanges in mammalian lungs}

\author[1,*,+]{Benoit Haut}
\author[1,+]{Cyril Karamaoun}
\author[2,+]{Benjamin Mauroy}
\author[1,3,+]{Benjamin Sobac}
\affil[1]{Transfers, Interfaces and Processes (TIPs), École polytechnique de Bruxelles, Université libre de Bruxelles, Brussels, Belgium}
\affil[2]{Laboratoire J.A. Dieudonné (UMR 7351), CNRS, Université de Nice Sophia-Antipolis, Nice, France}
\affil[3]{Laboratoire des Fluides Complexes et leurs Réservoirs (UMR 5150), CNRS, Université de Pau et des Pays de l’Adour, Anglet, France}

\affil[*]{Benoit.Haut@ulb.be}

\affil[+]{These authors contributed equally to this work. They appear in alphabetic order.}

\keywords{Heat and mass transfer, Lungs, Mammals, Maximal effort, Rest, Scaling laws}

\begin{abstract}
A secondary function of the respiratory system of the mammals is, during inspiration, to heat the air to body temperature and to saturate it with water before it reaches the alveoli. Relying on a mathematical model, we propose a comprehensive analysis of this function, considering all the terrestrial mammals (spanning six orders of magnitude of the body mass, M) and focusing on the sole contribution of the lungs to this air conditioning. The results highlight significant differences between the small and the large mammals, as well as between rest and effort, regarding the spatial distribution of heat and water exchanges in the lungs, and also in terms of regime of mass transfer taking place in the lumen of the airways. Interestingly, the results show that the mammalian lungs appear to be designed just right to fully condition the air at maximal effort (and clearly over-designed at rest, except for the smallest mammals): all generations of the bronchial region of the lungs are mobilized for this purpose, with calculated values of the local evaporation rate of water from the bronchial mucosa that can be very close to the maximal ability of the serous cells to replenish this mucosa with water. For mammals with a mass above a certain threshold ($\simeq 5$ kg at rest and $\simeq 50$ g at maximal effort), it appears that the maximal value of this evaporation rate scales as $M^{-1/8}$ at rest and $M^{-1/16}$ at maximal effort and that around 40\% (at rest) or 50\% (at maximal effort) of the water/heat extracted from the lungs during inspiration is returned to the bronchial mucosa during expiration, independently of the mass, due to a subtle coupling between different phenomena. This last result implies that, above these thresholds, the amounts of water and heat extracted from the lungs by the ventilation scale with the mass such as the ventilation rate (i.e. as $M^{3/4}$ at rest and $M^{7/8}$ at maximal effort). Finally, it is worth to mention that these amounts appear to remain limited, but not negligible, when compared to relevant global quantities, even at maximal effort (4-6\%).
\end{abstract}

\begin{document}

\flushbottom
\maketitle
\thispagestyle{empty}

\section*{Introduction}

In mammals, the respiratory system is divided into an upper and a lower tract. The upper tract includes the nose, the nasal cavities, the pharynx and the larynx, while the lower tract is composed of the lungs. The lungs form a dichotomous branching tree in which each level of subdivision is called a generation \cite{weibel_morphometry_1963} (Figures \ref{general}.a and \ref{general}.b). The lungs are themselves divided in two regions: the bronchial region, composed of airways (trachea, bronchi, bronchioles...), and the alveolar region. 

Besides the exchange of oxygen and carbon dioxide with the environment, an important role of the respiratory tract is, during inspiration, to heat the air to body temperature and to saturate it with water, before it reaches the alveoli. This is physiologically crucial because it protects the alveolar membrane from thermal injury and because oxygen and carbon dioxide cannot be exchanged if this membrane is dry \cite{Negus,Sherwood_2012,walker_heat_1961}. As sketched in Figure \ref{general}.d, for this "conditioning" to take place, heat is extracted from the wall of the airways to, on one hand, heat the air in the lumen and, on the other hand, evaporate water contained in the airway surface liquid (ASL)\cite{walker_heat_1961,Williams_1996,wu_numerical_2014}, a thin liquid layer covering the respiratory epithelium. The water in the ASL is secreted by serous cells in this epithelium and in submucosal glands, which are known to be abundant in the first generations of the lungs\cite{Joo_2006,Widdi}. The ASL also has the function of trapping the inhaled particles and pathogens and thus contributes to their removal by the mechanism of mucociliary clearance \cite{button_periciliary_2012,Chateau_2019,king_physiology_2006}. 

As heat is extracted from the wall of the airways during inspiration, the respiratory epithelium and the surrounding tissues are cooled down, especially in the upper tract and in the proximal generations of the lungs. Therefore, during expiration, the exhaled gas, which is fully humidified at body temperature when it leaves the alveolar region of the lungs, passes over cooler airway walls. Consequently, the temperature of this exhaled gas is reduced and it returns some of its water vapor to the ASL, by the process of condensation (Figures \ref{general}.c and \ref{general}.d) \cite{Kahn,Mcfadden_thermal_1985,Mcfadden_1992,Sherwood_2012,walker_heat_1961}.

Usually, the conditioning of the air is not complete in the upper respiratory tract and is thus still taking place in the first generation of the lungs \cite{Mcfadden_thermal_1985,Mcfadden_1992,walker_heat_1961}. McFadden et al. have measured that, when a human adult at rest breathes in a room with an air at a temperature of 27$^{\circ}$C and a relative humidity of 40\%, the temperature of the air at the top of its trachea during inspiration is approximately 33$^{\circ}$C and the air is not yet at the body temperature in the 6$^{\text{th}}$ generation of his lungs\cite{Mcfadden_thermal_1985}. 

Numerous works, notably based on the development of mathematical models, have been devoted to a detailed analysis of the water and heat exchanges taking place in the human lungs\cite{daviskas_mathematical_1990,Ferron_1985,ingenito_finite_1986,karamaoun_new_2018,Saidel1983,hanna_theoretical_1986,tawhai_modeling_2004,tsai_radial_1990,Tsu1988,warren_role_2010,wu_numerical_2014}. However, to the best of our knowledge, a generalization of the analysis to the whole class of mammals has never been carried out, although it could bring rich insights. 
In particular, the use of the allometric approach (i.e. the study of the relationship of body size to shape, anatomy and physiology) has gained importance in the last decades and can lead to strong results in terms of physiology and ecology \cite{west_general_1997}. With regards to the mammalian lungs, it is well known that several of their characteristics can be finely related to the mass of the body by allometric scaling laws \cite{Noel2022,Weibel2005,west_general_1997} 

The objective of this work is to gain insights into the water and heat exchanges in the lungs of the terrestrial mammals. Our methodology is based on the development of a comprehensive mathematical model of these exchanges, incorporating already known allometric scaling laws. This model extends one of our previous works, limited to the human species \cite{Haut2021}, to the whole class of terrestrial mammals, typically from a small mouse ($5$~g) to an elephant ($5000$~kg), and this in two situations: at rest and at maximum effort. With this model, we intend to analyze if there are notable differences in the mechanisms involved depending on the mass of the mammal, to identify the key parameters governing these water and heat exchanges and to derive scaling laws that cover the wide spectrum of the terrestrial mammal mass. In particular, we aim to characterize the dependence on the mass of the amounts of water and heat extracted from the lungs by the ventilation and of the evaporation rate of water from the ASL.

\section*{Mathematical formulation}

The allometric laws used in the model are summarized in Table \ref{allometriclaws} and the values of the parameters independent of the mass of the body are given in Table \ref{parametervalues}. All the numerical results provided in this paper have been obtained with these data. Moreover, additional information related to the development of the model is provided in the Supplementary Material (SM). In the figures provided hereafter and in the SM, "Mouse" corresponds to a mass $M = 5$ g, "Weasel" to $M = 50$ g, "Rat" to $M = 500$ g, "Cat" to $M = 5$ kg, "Human" to $M = 50$ kg, "Horse" to $M = 500$ kg and "Elephant" to $M =$ 5000 kg.

\subsection*{Geometrical representation of mammalian lungs}\label{definitions}

To represent the geometry of the bronchial region of mammalian lungs, we use the classical "Weibel A" model\cite{west_general_1997} (Figure \ref{general}.b). 
In this framework, the airways are considered as right circular cylinders and the bronchial region is represented as a dichotomous branching tree with, for a given mass $M$, the airways belonging to a given generation having all the same radius and diameter \cite{weibel_morphometry_1963}. For a given value of $M$, the ratio of the length to the radius of the airways is a constant. This aspect ratio, written $\beta$, is assumed to scale with the mass as \cite{west_general_1997} $\beta = \beta_{\text{ref}}(M/M_{\text{ref}})^{-1/8}$, with $\beta_{\text{ref}}$ the value of $\beta$ for a mammal with a reference mass $M_{\text{ref}}$. The generations in the bronchial region are numbered from $1$ to $n$, $1$ being the trachea and $n$ corresponding to the airways connected to the alveolar region (Figure \ref{general}.b). The number of airways in generation $i$ is thus $2^{i-1}$. 
The dimensions of the airways are considered constant during a respiratory cycle. This assumption is supported by the fact that the airways contributing mainly to water and heat exchanges are the proximal ones and that the dimensions of these airways experience little variations during a respiratory cycle, due to the significant presence of cartilage around them \cite{Haverkamp_2005}. The radius and length of the airways in generation $i$ are written $R_i$ and $L_i$, respectively (Figure \ref{general}.d), with thus $L_i = \beta R_i$. 
The radius of the trachea {$R_1$ is assumed to scale with the mass as \cite{west_general_1997} $R_1 = R_{1,\text{ref}}(M/M_{\text{ref}})^{3/8}$ and the following classical recurrence relation {is considered\cite{Mauroy2004,Noel2019,west_general_1997}: $R_i = h R_{i-1}$, with $h$ a constant independent of the mass of the body and equal to $2^{-1/3}$ (Figure \ref{general}.b). 
The number of generations in the bronchial tree $n$ is calculated by assuming that this tree stops when the successive divisions generate airways with a radius smaller than the alveolar diameter, $d_{\text{alv}}$ (i.e. $R_{n} < d_{\text{alv}}$ and $R_{n-1} > d_{\text{alv}}$). $d_\text{alv}$ is assumed to scale with the mass as \cite{west_general_1997} $d_\text{alv} = d_\text{alv,ref}(M/M_{\text{ref}})^{1/12}$. 

\subsection*{Transport of water in the lumen of the airways}

We assume that the air in the lungs is incompressible and that Fick's law for binary mixtures can be used to describe the diffusion of water vapor in air. We also assume that, on both inspiration and expiration, momentum and water vapor transports are stationary and present an axial symmetry in the lumen of the airways \cite{Haut2021}. 

We consider that the durations of inspiration and expiration are equal, as results obtained with a model developed previously show low sensitivity to the ratio of these durations\cite{Haut2021}. Consequently, the inspiration and expiration volumetric flow rates are equal; they are written $Q$. To express $Q$, we assume that, at rest, it is proportional to the basal metabolic rate (BMR) and that the latter scales with the mass as \cite{Kleiber,west_general_1997} $\text{BMR} \propto M^{3/4}$. Moreover, we assume that, at maximal effort, $Q$ is proportional to the maximum metabolic rate (MMR) and that the latter scales with the mass as \cite{painter,Weibel2005} $\text{MMR} \propto M^{7/8}$. Therefore, we can write $Q = \psi Q_{\text{ref}}(M/M_{\text{ref}})^{3/4}$, with $Q_{\text{ref}}$ the inspiration flow rate at rest for a mammal of mass $M_{\text{ref}}$ and $\psi \geq 1$ a factor accounting for the increase of the inspiration flow rate during a possible effort. $\psi = 1$ at rest and $\psi = \psi_{\text{ref}}\left(M/M_{\text{ref}}\right)^{1/8}$ at maximal effort, with thus $\psi_{\text{ref}}$ the value of $\psi$ at maximal effort for a mammal of mass $M_{\text{ref}}$.\\ 

Now, let us consider an airway in generation $i$ of the bronchial region of the lungs of a mammal with a given mass $M$, during either inspiration or expiration. $U_i = Q/(2^{i-1}\pi R_i^2)$ is the average axial velocity of the air in this airway. The P\'eclet number of the flow in the airway, written Pe$_i$, is defined as Pe$_i = L_i U_i/\mathcal{D}$, with $\mathcal{D}$ the diffusion coefficient of water in air. It compares a characteristic time of convective axial transport in the airway ($L_i/U_i$) and a characteristic time of axial diffusive transport ($L_i^2/\mathcal{D}$). Pe$_i  \gg 1$ means that axial convective transport dominates diffusive axial transport, while Pe$_i  \ll 1$ means the opposite. As the flow rate is conserved between generation $i$ and $i-1$, we can write that $2 \pi R_i^2 U_i = \pi R_{i-1}^2 U_{i-1}$. Hence, keeping in mind that $L_i/R_i = L_{i-1}/R_{i-1} = \beta$ and that $R_i = h R_{i-1}$, we have:
\begin{equation}\label{Peclet}
\text{Pe}_i = \frac{\text{Pe}_{i-1}}{2h}= \frac{\text{Pe}_{1}}{(2h)^{i-1}}=\frac{1}{(2h)^{i-1}}\frac{\beta Q}{\pi R_1 \mathcal{D}}=\frac{\psi}{2^{\frac{2(i-1)}{3}}}\left(\frac{M}{M_{\text{ref}}}\right)^{\frac{1}{4}}\text{Pe}_{1,\text{ref}}
\end{equation}
with Pe$_{1,\text{ref}}= \beta_{\text{ref}}Q_{\text{ref}}/(\pi R_{1,\text{ref}}\mathcal{D})$ the P\'eclet number in the trachea at the reference mass $M_{\text{ref}}$.

Considering a range of mass from 5 g to 5000 kg, we can evaluate that, at maximal effort, the P\'eclet number in the airways of the bronchial region of the lungs is always larger than 10, whatever the mass of the mammal or the generation considered (see Figure S1.b in the SM). At rest, P\'eclet lower than 10 can be reached in distal generations (for mammals with a mass larger than 500 g, see Figure S1.a in the SM). However, the results obtained with our model show that, at rest, these generations do not contribute to air conditioning (i.e. the air is saturated with water and at body temperature before reaching these generations). Consequently, in the following, we neglect the axial diffusive transport in the airways. 

As shown in the results section, a key dimensionless number is actually Pe$_i/\beta^2$, which depends on $M$, $i$ and $\psi$ as follows:
\begin{equation}\label{Pecletbis}
\frac{\text{Pe}_i}{\beta^2} = \frac{\psi}{2^{\frac{2(i-1)}{3}}}\sqrt{\frac{M}{M_{\text{ref}}}}\frac{\text{Pe}_{1,\text{ref}}}{\beta_{\text{ref}}^2}
\end{equation}  
Figures S1.c and S1.d in the SM show Pe$_i/\beta^2$ as a function of the generation index $i$, for mammals of different sizes, at rest and at maximal effort. These figures and Equation \ref{Pecletbis} show that Pe$_i/\beta^2$ increases with the size of the airway (i.e. increases with the mass and decreases with an increase of the generation index) and with the intensity of an effort.\\ 

To write a dimensionless transport equation for the water vapor in the lumen of the airway, we divide the air velocity vector by $U_i$ and we introduce dimensionless axial ($z$) and radial ($r$) coordinates (see Figure \ref{general}.d). $z$ is oriented in the direction of the flow, with $z=0$ at the beginning of the airway and $z=1$ at its end. $r=0$ at the center of the airway and $r = 1$ at the lumen--ASL interface. Let us write $c_i(z,r)$ the water vapor concentration at position $(z,r)$ in the lumen. According to our different assumptions, $c_i$ satisfies the following dimensionless transport equation:
\begin{equation}
\label{mass_local}
\frac{1}{\beta}\frac{\partial}{\partial z}(v_z c_i)+ \frac{1}{r} \frac{\partial}{\partial r}(r v_r c_i) =  \frac{\beta}{\text{Pe}_i} \frac{1}{r}\frac{\partial}{\partial r}\left(r \frac{\partial c_i}{\partial r}\right)
\end{equation}
with $v_z$ and $v_r$ the dimensionless axial and radial components of the air velocity vector in the lumen of the airway.

As the flow rate $Q$ is the same on inspiration and expiration, so do $U_i$ and Pe$_i$. The velocity field $(v_z, v_r)$ is also the same on inspiration and expiration, because the $z$ axis is oriented in the direction of flow. Note that this field obviously also depends on the generation index $i$, but this is not explicitly stated in its notation.

Multiplying Equation \ref{mass_local} by $2 r$ and integrating from $r = 0$ to $r = 1$ gives, if we assume a no-slip condition at the lumen--ASL interface (i.e. $v_r(z,r=1)= 0$), and as axial symmetry implies $\partial c_i/\partial r\vert_{r=0} = 0$:
\begin{equation}
\frac{d\bar{c}_i}{dz}=\left. \frac{2 \beta^2}{\text{Pe}_i}\frac{\partial c_i}{\partial r}\right|_{r=1}
\end{equation}
with $\bar{c}_i(z)$ the velocity-average of $c_i$ at position $z$, defined as:
\begin{equation}
\bar{c}_i(z) = 2 \int_0^1c_i(z,r) v_z(z,r) r dr
\end{equation}

To model the exchange of water between the surface of the ASL and the lumen of the airway, we first consider that the temperature of the air in contact with the ASL, written $T_{\mu,i}$ (Figure \ref{general}.d), is independent of $z$ and is the same during inspiration and expiration \cite{Haut2021}. As discussed in Haut et al.\cite{Haut2021} and shown in Wu et al.\cite{wu_numerical_2014}, it results from the fact that, in a single airway, neither the renewal of blood by the circulation nor the extraction of heat to condition the air have the ability to significantly modify the temperature profile in the bronchial wall (along $z$ or with time). The temperature of the bronchial wall slightly decreases during inspiration and slightly increases during expiration, but this remains limited. Moreover, we assume that the air in contact with the ASL is at thermodynamic equilibrium with this liquid, which is considered as pure water with regard to this equilibrium\cite{wu_numerical_2014}. Hence, the water vapor concentration in the air in contact with the ASL (i.e. $c_i(z,1)$) is written $c_{\mu,i}$ (Figure \ref{general}.d) and is expressed as $c_{\mu,i} = c_{\text{sat}}(T_{\mu,i})$, with $c_{\text{sat}}(T)$ the saturation concentration of pure water in air at the temperature $T$.

With this in mind, let us now introduce the Sherwood number of the mass transfer between the ASL and the lumen of the airway, Sh$_i$, classically defined as:
\begin{equation}
\text{Sh}_i=\frac{1}{c_{\mu,i} - \bar{c}_i }\left. \frac{\partial c_i}{\partial r}\right|_{r=1} 
\end{equation}

The corresponding dimensional mass transfer coefficient between the ASL and the lumen is thus $k_i = \text{Sh}_i \mathcal{D}/R_i$. 

If Sh$_{i}$ is assumed independent of the axial coordinate (the validity of this assumption is discussed in Haut et al.\cite{Haut2021}), the two last equations can be combined to give:
\begin{equation}\label{integral1}
\log\left(\frac{c_{\mu,i}-\bar{c}_i(1)}{c_{\mu,i}-\bar{c}_i(0)}\right)=-\frac{2 \beta^2 \text{Sh}_i}{\text{Pe}_i}
\end{equation}

This equation can be rearranged as:
\begin{equation}\label{bilinsp}
\bar{c}_i(1) = \bar{c}_i(0)+\Gamma_i (c_{\mu,i}-\bar{c}_i(1) )
\end{equation}
with:
\begin{equation}\label{gamma}
\Gamma_i = \exp\left(\frac{2 \beta^2 \text{Sh}_i}{\text{Pe}_i}\right)-1
\end{equation}

$\Gamma_i$ is a dimensionless number characterizing the ability of the mass transfer phenomena in the airway to condition the air. In Equation \ref{bilinsp}, it multiplies a driving force, $c_{\mu,i} - \bar{c}_i(1)$, to express the increase of the water vapor concentration when air flows through the airway, $\bar{c}_i(1) - \bar{c}_i(0)$. $\Gamma_i \to \infty$ implies $\bar{c}_i(1) \to c_{\mu,i}$, a conditioning to the maximal possible extent, while $\Gamma_i \to 0$ implies $\bar{c}_i(1) \to \bar{c}_i(0)$, i.e. no conditioning. 

As fully detailed in Section 2 of the SM, numerical simulations of mass and momentum balance equations in the airway allow establishing the following correlation between Sh$_i$ and Pe$_i/\beta^2$ (see Figure S3 in the SM), which is valid for Pe$_i > 10$:
\begin{equation}\label{shappr}
\text{Sh}_i = 1.3 + \sqrt{\frac{\text{Pe}_i}{2\beta^2}}
\end{equation}


Let us now consider specifically the inspiration. We write $c_i^{\text{insp}}$ the velocity-averaged water vapor concentration at the distal extremity of an airway in generation $i$ during inspiration. Moreover, we write $c_0^{\text{insp}}$ the velocity-averaged water vapor concentration of the air entering the lungs during inspiration (assumed to be known, as a boundary condition of the problem). As there is a continuity of the water vapor concentration between successive generations, Equation \ref{bilinsp} can be rewritten as:
\begin{equation}\label{eqbil1}
c_i^{\text{insp}}  = c_{i-1}^{\text{insp}} +\Gamma_i (c_{\mu,i}-c_i^{\text{insp}} )
\end{equation}

Regarding expiration, we write $c_{i}^{\text{exp}}$ the velocity-averaged water concentration in the lumen at the distal extremity of generation $i$ during expiration (thus $c_{i}^{\text{exp}}$ and $c_{i}^{\text{insp}}$ measure concentrations at the same location). Moreover, we write $c_{0}^{\text{exp}}$  the velocity-averaged water vapor concentration in the air leaving the lungs during expiration (which is an unknown). An equation similar to Equation \ref{eqbil1} can be written straightforwardly for expiration:
\begin{equation}\label{eqbil2}
c_{i-1}^{\text{exp}} = c_{i}^{\text{exp}} +\Gamma_i (c_{\mu,i} - c_{i-1}^{\text{exp}} )
\end{equation}

\subsection*{Heat exchange between the bronchial wall and the ASL--lumen interface}

To model the heat exchange between the bronchial wall and the ASL--lumen interface, we first assume that the evaporation of water from the ASL is the sole cause of heat extraction from the tissues surrounding the bronchial epithelium. Consequently, in the following, we can write that the amount of heat extracted from the wall of an airway is equal the amount of water transferred from the ASL to the lumen of this airway, multiplied by the latent heat of vaporization of water. Results presented in a previous work\cite{Haut2021} show that this assumption does not prevent a very precise estimate of the water vapor concentration profiles in the lungs. It is supported by the fact that, except when breathing very cold air, the heat used to evaporate water usually accounts for 80--90\% of the total heat extracted from the lungs for the conditioning of the air \cite{Haut2021}. 

We introduce a characteristic time $t_{\text{bl}}$, defined as the volume of the vascularized tissues surrounding the bronchial epithelium (lamina propria, muscles...) divided by the flow rate of blood within them. We assume that it can be expressed as \cite{west_general_1997} $t_\text{bl} = (t_{\text{bl},\text{ref}}/\phi)(M/M_{\text{ref}})^{1/4}$, with $t_{\text{bl},\text{ref}}$ this characteristic time at rest for a mammal of mass $M_{\text{ref}}$ and $\phi \geq 1$ a factor accounting for the increase of the cardiac output during a possible effort. $\phi = 1$ at rest and we assume $\phi = \phi_{\text{ref}}(M/M_{\text{ref}})^{1/8}$ at maximal effort (i.e. we assume the same dependence of $\phi$ and $\psi$ on the mass at maximal effort). 
For a human adult at rest, the ratio of the volume of blood in the body to the cardiac output is of the order of the minute. Moreover, the volume fraction of blood in a tissue is typically 2--4\% \cite{Hinde_2017}. Thus, at rest and for a human adult, $t_\text{bl}$ is around 1500-3000 s.

The tissues surrounding the bronchial epithelium are impacted by heat transfer on a length scale $\delta_w = \sqrt{\alpha_w t_{\text{bl}}}$, sketched in Figure \ref{general}.d, with $\alpha_w$ the thermal diffusivity of these tissues 
(assumed equal to the one of pure water\cite{warren_role_2010}). It can be evaluated that this length scale is way larger than the thickness of the ASL and of the epithelium (both having a thickness of approximately 10 $\mu$m\cite{Jeffery_1998,Widdi}). Therefore, an estimation of the average, over a whole respiratory cycle, heat flux from the tissues to the ASL--lumen interface in generation $i$ is $\lambda_w (T_b-T_{\mu,i})/ \sqrt{\alpha_w t_{\text{bl}}}$, with $\lambda_w$ the thermal conductivity of the tissues surrounding the bronchial epithelium (also assumed equal to the one of pure water \cite{warren_role_2010}). Consequently, we can propose the following equation to describe the heat exchange between the bronchial wall and the ASL--lumen interface:

\begin{equation}\label{interf1}
\lambda_w \frac{T_b-T_{\mu,i}}{\sqrt{\alpha_w t_{\text{bl}}}} = \frac{\text{Sh}_i \mathcal{D}}{R_i}(c_{\mu,i} - [c]_i)\mathcal{L}
\end{equation}
with $\mathcal{L}$ the molar heat of vaporization of water and:
\begin{equation}
[c]_i = \frac{c_{i}^{\text{insp}}+c_{i-1}^{\text{insp}}+c_{i}^{\text{exp}}+c_{i-1}^{\text{exp}}}{4}
\end{equation}
an approximation of the time and space average of the water vapor concentration in generation $i$.\\

Moreover, the equation describing the thermodynamic equilibrium at the ASL--lumen interface can be linearized:
\begin{equation}\label{lineq}
c_{\mu,i} = c_{\text{sat}}(T_{\mu,i}) \simeq c_{\text{sat}}(T_b) + \left . \frac{dc_{\text{sat}}}{dT}\right|_{T =T_b}(T_{\mu,i}-T_b)
\end{equation}

Eliminating $T_{\mu,i}$ between Equations \ref{interf1} and \ref{lineq} gives:
\begin{equation}\label{eqint}
\Lambda_i (c_{\text{sat}}(T_b) - c_{\mu,i}) = c_{\mu,i} - [c]_i
\end{equation}
with:
\begin{equation}\label{lambda}
\Lambda_i = \frac{\sqrt{\phi}\lambda_w R_{1,\text{ref}}}{2^{\frac{i-1}{3}}\mathcal{L}\sqrt{ \alpha_w t_{\text{bl,ref}}}\left . \frac{dc_{\text{sat}}}{dT}\right|_{T =T_b}\text{Sh}_i \mathcal{D}}\left(\frac{M}{M_{\text{ref}}}\right)^{\frac{1}{4}}
\end{equation}

$\Lambda_i$ is a dimensionless number measuring, in generation $i$, the ability of the vascularization to heat the tissues surrounding the bronchial epithelium, in front of the intensity of heat extraction from these tissues to evaporate water in the ASL. $\Lambda_i \gg 1$ means that heat supply by the vascularization dominates heat extraction (and in this case Equation \ref{eqint} shows that $c_{\mu,i} \simeq c_{\text{sat}}(T_b)$), while $\Lambda_i \ll 1$ means the opposite. 

\subsection*{Model summary}

Our model of the water and heat exchanges in the bronchial region of the mammalian lungs is composed of Equations \ref{eqbil1} and \ref{eqbil2}, completed by Equation \ref{eqint}, the three of them written for $i = 1, ..., n$. The two key dimensionless numbers characterizing these exchanges, $\Gamma_i$ and $\Lambda_i$, can be calculated, for a given mass $M$ and given values of $\phi$ and $\psi$, using Equations \ref{Pecletbis}, \ref{gamma}, \ref{shappr} and \ref{lambda}.  
The model includes $3n$ unknowns: $c_{\mu,i} (\forall i = 1, ..., n$), $c_{i}^{\text{insp}} (\forall i = 1, ..., n$) and $c_{i}^{\text{exp}} (\forall i = 0, ..., n-1$). $c_0^{\text{insp}}$ is a boundary condition of the problem and $c_{n}^{\text{exp}}$, the water vapor concentration in the air entering the bronchial region of the lungs during expiration (i.e. leaving the alveolar region), is assumed equal to $c_{\text{sat}}(T_b)$. This assumption is well-supported by the results obtained with our model since, in all cases considered, the air is fully conditioned before reaching the alveolar region during inspiration.

The model can be made dimensionless by introducing dimensionless water vapor concentrations as follows. To  
a given dimensional concentration $c$, we associate the dimensionless concentration $\tilde{c}$ given by:
\begin{equation}\label{adimc}
\tilde{c}= \frac{c-c_{0}^{\text{insp}}}{c_{\text{sat}}(T_b)-c_{0}^{\text{insp}}}
\end{equation} 
This dimensionless concentration corresponds to a measure of the extent of the air conditioning, as $\tilde{c} = 0$ means $c = c_{0}^{\text{insp}}$ and $\tilde{c} = 1$ means $c = c_{\text{sat}}(T_b)$ (i.e. the air is saturated with water and at body temperature). In its dimensionless form, the model has the following boundary conditions: $\tilde{c}_0^{\text{insp}} = 0$ and $\tilde{c}_n^{\text{exp}} = 1$. This dimensionless model is thus independent of $c_0^{\text{insp}}$ (i.e. on the atmospheric conditions), as $c_0^{\text{insp}}$ does not appear in the expressions of the dimensionless numbers $\Gamma_i$ and $\Lambda_i$. Consequently, all the dimensionless quantities derived from the solving of the model are also independent on the atmospheric conditions.


\section*{Results and discussion}


\subsection*{Key dimensionless numbers}

Let us begin by presenting the distribution of the two key dimensionless numbers characterizing locally the water and heat transfers in the bronchial tree of mammals, $\Gamma_i$ (Equation \ref{gamma}) and $\Lambda_i$ (Equation \ref{lambda}). Figure \ref{fig2} shows these two dimensionless numbers as functions of the generation index $i$ and for various mammal mass $M$, at rest (left column) and at maximal effort (right column), respectively. 

Figure \ref{fig2}.a-b shows that $\Gamma_i$, which characterizes the ability of the mass transfer phenomena in the lumen to condition the air, increases with the generation index $i$, decreases with an increase of the mass of the mammal and decreases in case of an effort (please note the significant difference in scale between the $y$ axes of Figures \ref{fig2}.a and \ref{fig2}.b). 
Figure \ref{fig2} also evidences that, while proximal airways of the large mammals appears to be "poor conditioners", especially at maximal effort, since $\Gamma_i \lesssim 1$, the distal airways are "super conditioners", especially at rest, since $\Gamma_i \gg 1$. This difference in behavior can be understood by analyzing how the Sherwood number Sh$_i$ depends on the size of the airways and on the intensity of an effort. 

On the one hand, for Pe$_i/\beta^2 \gg 1$, i.e. for large airways and increasing effort, Sh$_i \simeq \sqrt{\text{Pe}_i/(2\beta^2)}$ (see Equation \ref{shappr}). This 1/2 exponent is typical from a "boundary layer" regime of mass transfer at a surface with a "plug flow" over it \cite{Clift_1978,Schlichting_2000}. In our situation, "boundary layer" regime means that the diffusion boundary layer remains located near the ASL--lumen interface in the entire airway, as sketched in Figure \ref{general}.d and also presented in Figure S2.c in the SM. Thus, from Equation \ref{gamma}, a limit expression of $\Gamma_i$ can be obtained for Pe$_i/\beta^2 \gg 1$:

\begin{equation}\label{gammalim}
\Gamma_i \simeq \Gamma_{i,\text{lim}} = \sqrt{\frac{2 \beta^2}{\text{Pe}_i}}
\end{equation}

On the other hand, for Pe$_i/\beta^2$ between 0.1 and 10, i.e. for small airways and decreasing effort, Sh$_i\simeq1.5$  (see Equation \ref{shappr}). In this "constant Sherwood" regime of mass transfer, the boundary layer that develops on the ASL--lumen interface eventually fills the entire airway, as shown in Figures S2.a and S2.b in the SM. Thus, in a given airway, conditioning is almost carried out to the maximum possible extent (i.e. the air leaving the airway is at thermal and mass equilibrium with the ASL, $\bar{c}_i(1) \simeq c_{\mu,i}$); the temperature of the ASL in this airway is established such as to obtain an equilibrium between the amount of heat extracted from the tissues and brought to the ASL--lumen interface and the amount of heat needed to saturate the air with water at the temperature of the ASL. Thus, from Equation \ref{gamma}, an approximate expression of $\Gamma_i$ can be obtained for small values of Pe$_i/\beta^2$:

\begin{equation}\label{gammasmall}
\Gamma_i \simeq   \Gamma_{i,\text{sml}}= \exp\left(\frac{3 \beta^2}{\text{Pe}_i}\right)-1 
\end{equation}

Interestingly, this last equation, combined with Equation \ref{Pecletbis}, enables to evidence that, for small values of Pe$_i/\beta^2$, $\Gamma_i$ increases faster than exponentially with $i$, explaining mathematically why the small airways are "super-conditioners". Physically, it is the result of the large surface area to volume ratio and of the division of the flow into a very large number of conducts. 

In Figure \ref{fig2}.a-b, $\Gamma_{i,\text{lim}}$ and $\Gamma_{i,\text{sml}}$ are given, but only for half of the mammals for the sake of clarity. We 
see that, while the small mammals only experience the "constant Sherwood" regime of mass transfer in the bronchial region of their lungs, especially at rest, there is, for the large mammals, a transition between the "boundary layer" and the "constant Sherwood" regimes, especially at maximal effort.\\

Figure \ref{fig2}.c-d shows that $\Lambda_i$, which characterizes the ability of the vascularization to maintain the bronchial wall at the body temperature, decreases with an increase of the generation index $i$, increases with the mass of the mammal and decreases in the case of an effort. $\Lambda_i$ is significantly smaller than 1 for all mammals, whatever the generation or the intensity of the effort. 

As it depends also on Sh$_i$ (see Equation \ref{lambda}), $\Lambda_i$ also shows a change of regime between small and large values of Pe$_i/\beta^2$. On the one hand, since Sh$_i \simeq \sqrt{\text{Pe}_i/(2\beta^2)}$ for Pe$_i/\beta^2 \gg 1$, a limit expression of $\Lambda_i$ can be obtained from Equations \ref{Pecletbis} and \ref{lambda} when Pe$_i/\beta^2 \gg 1$: 
\begin{equation}\label{lambdalim}
\Lambda_i \simeq \Lambda_{\text{lim}} =  \Theta \sqrt{\frac{\phi}{\psi}}
\end{equation}
with:
\begin{equation}\label{analtheta}
\Theta = \frac{\lambda_w R_{1,\text{ref}}^{3/2}}{\mathcal{L}\sqrt{\alpha_w t_{\text{bl,ref}}}\left . \frac{dc_{\text{sat}}}{dT}\right|_{T =T_b} }\sqrt{\frac{2\pi \beta_{\text{ref}}}{\mathcal{D}Q_{\text{ref}}}}
\end{equation}

$\Theta$ is a constant, equal to 0.37. We see that $\Lambda_i$ is independent of the generation index and of the mass. It is proportional to $\sqrt{\phi/\psi}$, which is equal to 1 at rest and to 0.5 at maximal effort (see Table \ref{allometriclaws}). This explains why $\Lambda_i$ is smaller at maximal effort than at rest: the heating of the mucosa by the vascularization is disadvantaged compared to the extraction of heat from it by the evaporation. Consequently, air cooling and water condensation on expiration are expected to be favored when an effort is made. In Figure \ref{fig2}.c-d, the horizontal dashed lines provide the values of $\Lambda_{\text{lim}}$ at rest and at maximal effort.

On the other hand, as mentioned previously, for Pe$_i/\beta^2$ between 0.1 and 10, Sh$_i\simeq1.5$. Thus, from Equation \ref{lambda}, an approximate expression of $\Lambda_i$ can be obtained for small values of Pe$_i/\beta^2$:
\begin{equation}\label{lambdasmall}
\Lambda_i \simeq \Lambda_{i,\text{sml}} = \frac{\sqrt{\phi}\lambda_w R_{1,\text{ref}}}{2^{\frac{i-1}{3}}\mathcal{L}\sqrt{ \alpha_w t_{\text{bl,ref}}}\left . \frac{dc_{\text{sat}}}{dT}\right|_{T =T_b}\frac{3}{2}\mathcal{D}}\left(\frac{M}{M_{\text{ref}}}\right)^{\frac{1}{4}}
\end{equation}
This equation shows that, for small values of Pe$_i/\beta^2$, $\Lambda_i$ decreases with an increase of the generation index $i$ and increases with the mass (please remind that $\phi = 1$ at rest and $\phi \propto M^{1/8}$ at maximal effort), thus suggesting a larger decrease in the bronchial temperature of the small airways due to the evaporative cooling. In Figure \ref{fig2}.c-d, the dotted lines provide values of $\Lambda_{i,\text{sml}}$.\\

\subsection*{Water vapor concentration profiles in the lumen of the lungs}

Now, we can analyze the consequence of these local transfers on the distribution of the water vapor concentration along the bronchial region of mammalian lungs. For this purpose, Figure \ref{fig3} presents the dimensionless water vapor concentration profiles in the lumen of the lungs, during inspiration (full circles) and expiration (triangles), for mammals of different sizes, at rest (a) and at maximal effort (b). The empty circles give to the water concentration in the air at the lumen--ASL interface $\tilde{c}_{\mu,i}$. As a reminder, these dimensionless profiles do not depend on the atmospheric conditions. 

Figure \ref{fig3} shows that, on inspiration, the air is fully conditioned before leaving the bronchial region of the lungs in all cases considered. Moreover, on expiration, a significant condensation of water occurs, as the dimensionless water vapor concentration at the top of the trachea during expiration $\tilde{c}_0^{\text{exp}}$ is between 0.4 and 0.7, whatever the mass or the intensity of an effort. These two features directly result from  the "super-conditionner" property of the small airways, characterized by $\Gamma_i\gg1$, and from the poor ability of the bronchial wall to remain isothermal at body temperature since  $\Lambda_i <1$, respectively. Figure \ref{fig3}.b also shows that, when exercising, the decrease in $\Gamma_i$ and $\Lambda_i$ implies that more generations are needed for the conditioning. At maximal effort, the first generations of large mammals hardly contribute to air conditioning since they are "poor conditioners". Finally, it is striking to note that, at maximal effort and whatever the mammal, all the generations of the bronchial region of the lungs are mobilized to condition the air before it enters the alveolar region.\\

\subsection*{Distribution of the water and heat exchanges in the lungs}
We pursue by analyzing how the water and heat exchanges are distributed along the bronchial region of the mammalian lungs. To that end, we evaluate from the solution of the model the relative contribution of generation $i$ to these exchanges, by comparing the amount of water extracted per unit of time from the ASL of all the airways in this generation:
\begin{equation}
E_{w,i} = \frac{Q}{2}(c_i^{\text{insp}}-c_{i-1}^{\text{insp}} + c_{i-1}^{\text{exp}}-c_{i}^{\text{exp}})
\end{equation}
to the total amount of water extracted per unit of time from the lungs: 
\begin{equation}\label{totwextr}
E_w = \sum_{i = 1}^{i = n}E_{w,i} = \frac{Q}{2}( c_0^{\text{exp}}-c_0^{\text{insp}}  )
\end{equation}
To obtain the last equation, we consider that $c_n^{\text{insp}} = c_n^{\text{exp}} = c_{\text{sat}}(T_b)$, as the results show that, during inspiration, the air is fully conditioned before reaching the alveolar region in all the cases considered. 

The amount of heat extracted per unit of time from generation $i$ being $E_{h,i} = \mathcal{L}E_{w,i}$ and the total amount of heat extracted per unit of time from the lungs being $E_h= \mathcal{L}E_w$, it is worth to realize that $E_{w,i}/E_w = E_{h,i}/E_h$. 

Figure \ref{fig4} shows the distribution of the ratio $E_{w,i}/E_w$ along the bronchial tree for mammals of different sizes, at rest (a) and at maximal effort (b). 
For the small mammals, we see that $E_{w,i}/E_w$ monotonously decreases along the bronchial tree 
while, for large mammals, it first increases, before attaining a maximal value, and then decreases to zero when reaching the alveolar region. Moreover, for these large mammals, we see that the generation in which the largest amount of water is extracted gradually progresses into the bronchial tree with an increase of the mammal mass. This behavior can be understood by looking at Equation \ref{eqbil1}: it shows that the increase in water vapor concentration when the air flows through generation $i$ during inspiration, which is proportional to the amount of water evaporated in this generation during inspiration, is the product of a driving force, $c_{\mu,i}-c_i^{\text{insp}} $, tending to zero as $i$ increases, and of $\Gamma_i$, increasing with $i$ but tending to 0 as $M$ increases (see Equations \ref{Pecletbis}, \ref{gamma} and \ref{shappr}). Therefore, there is necessarily a threshold value of $M$ such that, when the mass of a mammal is above this threshold, it is in a generation of its lungs other than the trachea that the largest amount of water is extracted per unit of time. 

In addition, Figure \ref{fig5} provides the index of the generation in which the largest amount of water is extracted, written $i_{\text{max}}$, as a function of the mammal mass, at rest and at maximal effort. This figure evidences again that the generation in which the largest amount of water (or heat) is extracted is the trachea as long as the mass is below a certain threshold, which differs at rest and at maximal effort: it is around 5 kg (a cat) at rest and around 50 g (a weasel) at maximal effort. Then, when the mass is above this threshold, the generation in which the largest amount of water (or heat) is extracted gradually progresses into the bronchial tree with an increase of the mass. 

In Figure \ref{fig5}, the results of the model are compared with the prediction of the following analytical approximation (dashed lines), whose derivation is detailed in Section 3 of the SM:

\begin{equation}\label{eqanalimax}
i_{\text{max}} \simeq i_{\text{max}}^* = \max\left(1,a+ \frac{1.5\log(M)+ 3\log(\psi)}{\log(2)}\right)
\end{equation}
with $a$ a constant and $\max(x,y)$ the maximum of $x$ and $y$. Let us just specify here that, to obtain this equation, condensation during expiration is neglected and $\Gamma_i$ is taken equal to $\Gamma_{i,\text{lim}}$ (Equation \ref{gammalim}), which is a valid approximation when Pe$_i/\beta^2 \gg 1$. 

Interestingly, we see that this analytical approximation predicts very well how $i_\text{max}$ progresses into the bronchial tree when the inspiration flow rate or the mass increases. Notably, it evidences that 
$i_{\text{max}}$ increases by $3\log(\psi)/\log(2)$ when an effort is realized. 


\subsection*{Effectivity of water and heat extraction from the lungs}

To give further insights into the complexity of the water and heat exchanges in the mammalian lungs, it is interesting to consider local and overall effectivities of extraction. The local effectivity of water extraction in generation $i$ is defined as:
\begin{equation}\label{effloc}
\eta_i = \frac{c_i^{\text{insp}}-c_{i-1}^{\text{insp}}+c_{i-1}^{\text{exp}}-c_{i}^{\text{exp}}}{c_i^{\text{insp}}-c_{i-1}^{\text{insp}}}
\end{equation}
and corresponds to the ratio of the amount of water extracted from the bronchial wall in generation $i$ during a whole respiratory cycle to the amount of water extracted in this generation during inspiration only. As the bronchial wall is below the body temperature, condensation occurs during expiration (i.e. $c_{i-1}^{\text{exp}}<c_{i}^{\text{exp}}$) and $\eta_i < 1$. $\eta_i \simeq 1$ means that almost no condensation occurs during expiration (i.e. $c_{i-1}^{\text{exp}}\simeq c_{i}^{\text{exp}}\simeq c_{\text{sat}}(T_b)$). At the opposite, $\eta_i \simeq 0$ means that the water extracted from the ASL during inspiration is almost totally condensed during expiration. 

The overall effectivity of water extraction from the lungs, written $\eta,$ is defined as the ratio of the total amount of water extracted per unit of time from the lungs $E_w$ (Equation \ref{totwextr}) to its maximal possible value, obtained for $c_0^{\text{exp}} = c_{\text{sat}}(T_b)$, which corresponds to air leaving the lungs while being saturated with water and at body temperature. Hence:
\begin{equation}\label{effglob}
\eta = \frac{c_0^{\text{exp}}-c_0^{\text{insp}}}{c_{\text{sat}}(T_b)-c_0^{\text{insp}}} = \tilde{c}_0^{\text{exp}}
\end{equation}

$\eta$ appears equal to the dimensionless water vapor concentration at the top of the trachea during expiration, $\tilde{c}_0^{\text{exp}}$ (see Equation \ref{adimc}). Moreover, as the air is fully conditioned during inspiration, $1-\eta$ is the ratio of the amount of water condensed during expiration to the amount of water evaporated during inspiration. 

Please note that, even though $\eta_i$ and $\eta$ are defined as effectivities of water extraction from the lungs by the ventilation, they also measure effectivities of heat extraction, as water and heat removals are proportional to each other. 

Figure \ref{fig6} presents local and global effectivities of extraction as functions of the mammal mass, at rest (Figure \ref{fig6}.a) or at maximal effort (Figure \ref{fig6}.b). For a given mammal mass, three effectivities are presented: the local effectivity in the trachea $\eta_1$, the local effectivity in the generation in which the largest amount of water is extracted  (i.e. for $i = i_{\text{max}}$) $\eta_{i_{\text{max}}}$, and the overall effectivity of extraction $\eta$. As they are dimensionless, these effectivities are independent of the atmospheric conditions. Please note that the discontinuities in the plot of $\eta_{i_{\text{max}}}$ take place at the values of the mass for which $i_{\text{max}} $ increases by one unit.

Figure \ref{fig6} 
evidences a significant condensation on expiration, markedly limiting the hydric loss, as already observed in Figure \ref{fig3}. As mentioned previously, it is a consequence of the low values of $\Lambda_i$ in front of 1; this parameter measuring the ability of the vascularization to heat the tissues surrounding the bronchial epithelium. 
The results also show that, as expected, an effort leads to a decrease of the local or overall effectivity of extraction (i.e. to an increased proportion of water condensed on expiration, as already shown in Figure \ref{fig3}). This is linked to the decrease of $\Lambda_i$ when an effort is made (see Figure \ref{fig2}). 

Moreover, Figure \ref{fig6} shows a clear change of trend between the small and the large mammals, especially considering $\eta_{i_{\text{max}}}$. For the small mammals, $i_{\text{max}} = 1$ (see Figure \ref{fig5}) and $\eta_{i_{\text{max}}} (= \eta_1)$ increases with the mass. It is a consequence of the significant increase of $\Lambda_i$ with the mass for the small mammals (see Figure \ref{fig2}.c-d). Then, when the mass becomes larger than its threshold value beyond which $i_{\text{max}}$ is starting to progress along the bronchial tree (around 5 kg at rest and around 50 g at maximal effort), $\eta_{i_{\text{max}}}$ becomes almost independent of the mass. This is due to a complex interplay between several phenomena and, to understand this, let us consider for instance the case of maximal effort. Simultaneously looking at Figures \ref{fig2}.d and \ref{fig4}.b, we can observe that, although $\Lambda_i$ can vary strongly depending on $M$ and $i$, its value in the generation in which the largest amount of water is extracted  ($i = 1$ for the weasel, $i = 4$ for the rat, $i = 8$ for the cat, $i = 11$ for the human, $i = 14$ for the horse and $i = 17$ for the elephant) is almost constant (close to $0.11-0.12$), due to the progression of this generation down the bronchial tree when the mammal mass increases. And $\Lambda_i$ is the key parameter governing the balance, in generation $i$, between heat extraction from the tissues to evaporate water and heat supply by vascularization. 
Interestingly, we also observe that $\eta$ is very similar and close to $\eta_{i_{\text{max}}}$, which highlights that the overall effectivity is logically strongly linked to the processes taking place in the generation in which the largest amount of water is extracted. Consequently, except for small mammals, $\eta$ is quasi-insensitive to the mass, with a value close to 0.6 at rest and to 0.5 at maximal effort. Thus, approximately 40\% of the water evaporated on inspiration is condensed on expiration at rest, while 50\% is at maximal effort. 

Finally, as detailed in Section 4 of the SM and also shown in a previous work\cite{Sobac_2020}, the following analytical approximation for $\eta_i$ can be obtained, if $\Lambda_i$ is taken equal to $\Lambda_\text{lim}$ (Equation \ref{lambdalim}) and if some slight modifications are made on the model equations:

\begin{equation}\label{analrend}
\eta_i \simeq \eta^* =2 \sqrt{\Lambda_{\text{lim}} (1+\Lambda_{\text{lim}})}-2 \Lambda_{\text{lim}}= 2\Theta\left(\sqrt{\frac{1}{\Theta} \sqrt{\frac{\phi}{\psi}} + \frac{\phi}{\psi} }- \sqrt{\frac{\phi}{\psi}}\right)
\end{equation}
with $\Theta$ a constant, given by Equation \ref{analtheta}. Such as $\Lambda_{\text{lim}}$, $\eta^*$ appears independent of the mass at rest and at maximal effort. It is independent of the generation index and, thus, $\eta^*$ is also an approximation of the overall effectivity of extraction $\eta$. We calculate that, at rest, $\eta^* = 0.68$, and that, at maximal effort, $\eta^* = 0.57$. As shown in Figure \ref{fig6}, these analytical approximations of the effectivity (dashed horizontal lines) are close to those calculated with the model, except for small mammals (especially at rest), which is quite logical as Equation \ref{analrend} is obtained by taking $\Lambda_i = \Lambda_{\text{lim}}$ (i.e. in the limit Pe$_i/\beta^2 \gg 1$).\\

\subsection*{Total amounts of water and heat extracted from the lungs / evaporation rate from the ASL}

Let us end this results section by quantifying and analyzing the dependence with the mass of dimensional quantities of interest: the total amounts of water and heat extracted from the lungs by the ventilation and the evaporation rate of water from the ASL.

First, using the definition of the overall effectivity (Equation \ref{effglob}), the total amount of water extracted per unit of time from the lungs by the ventilation $E_w$ (Equation \ref{totwextr}) can be rewritten as follows:
\begin{equation}\label{totwextrbis}
E_w =\eta \frac{Q}{2}( c_{\text{sat}}(T_b)-c_0^{\text{insp}}  )
\end{equation}

Consequently, using Equation \ref{analrend}, an analytical approximation of $E_w$ can be simply derived:
\begin{equation}\label{totwextranal}
E_w \simeq E^*_w = \eta^* \frac{Q}{2}( c_{\text{sat}}(T_b)-c_0^{\text{insp}}  ) = \Theta\left(\sqrt{\frac{1}{\Theta} \sqrt{\frac{\phi}{\psi}} + \frac{\phi}{\psi} }- \sqrt{\frac{\phi}{\psi}}\right) Q (c_{\text{sat}}(T_b)-c_0^{\text{insp}}  )
\end{equation}
Regarding the total amount of heat extracted per unit of time from the lungs by the ventilation, we can thus use the following approximation: $E_h \simeq E^*_h = \mathcal{L} E^*_w$.

As $\eta^*$ is independent of the mass at rest or at maximal effort, we see with Equation \ref{totwextranal} that $E_w$ (and thus $E_h$) is expected to scale with the mass approximately as the inspiration flow rate $Q$ does (see Table \ref{allometriclaws}):
\begin{equation}
E_w \propto \left\{
    \begin{array}{ll}
        M^{3/4} & \mbox{ at rest,}  \\
        M^{7/8} & \mbox{ at maximal effort.}
    \end{array}
\right.
\end{equation}
Please note that these scalings may not be accurate for smaller mammals, since $\eta^*$ deviates from $\eta$ for them (see Figure \ref{fig6}). Note that the scaling with the mass at rest is also proposed in a previous work\cite{Sobac_2020}. 

$E_w$, $E_h$ and the evaporation rate from the ASL depend on the temperature and humidity at the top of the trachea during inspiration (through the parameter $c_0^{\text {insp}}$), themselves dependent on the atmospheric conditions and phenomena taking place in the upper respiratory tract. These atmospheric conditions can be very different from one environment to another and the morphology of the upper respiratory tract can also vary greatly from one species of mammal to another. Moreover, there is a switch from nose breathing at rest to a mixed nose-mouth breathing during an effort. These elements make it difficult to make a thorough analysis of the influence of the breathing conditions on $E_w$, $E_h$ and the evaporation rate from the ASL. Consequently, in this subsection, we limit ourselves to the use of a single value of $c_0^{\text {insp}}$, equal to half the saturation concentration of water in air at the body temperature: $c_0^{\text {insp}} = c_{\text{sat}}(T_b)/2 = 1.22$ mol m$^{-3}$. Thus, it allows considering a kind of "average" situation in terms of absolute humidity of the air at the entrance of the trachea. 

Figure \ref{fig7}.a presents the total amounts of water and heat extracted per unit of time from the lungs by the ventilation, $E_w$ and $E_h$ (please remind that $E_h = \mathcal{L}E_w$), as functions of the mass, at rest and at maximal effort. The analytical approximations $E^*_w$ and $E^*_h$ (Equation \ref{totwextranal}) are also presented for comparison. Figure \ref{fig7}.b presents the time average over a whole respiratory cycle of the evaporation rate in the trachea, written $J_1$, as a function of the mass, at rest and at maximal effort. $J_1$ is calculated as $J_1 = E_{w,1}/(2\pi R_1 L_1)$. Please note that the calculation results show that it is in the trachea that the evaporation rate is the largest, for all the cases considered. $E_w$ has been expressed in ml of liquid water per hour and $J_1$ has been expressed in $\mu$m$^3$ of liquid water per minute and per $\mu$m$^2$ of the ASL--lumen interface, using the volumetric mass of liquid water at the body temperature (whose value is given in Table \ref{parametervalues}). 

As expected, we see in Figure \ref{fig7}.a that, except for small mammals, $E_w$ and $E_h$ are close to $E_w^*$ and $E_h^*$. This confirms that $E_w$ and $E_h$ are almost proportional to $M^{3/4}$ at rest and to $M^{7/8}$ at maximal effort, such as the inspiration flow rate $Q$ does, because the overall effectivity $\eta$ is almost independent of the mass. 

The results presented in Figure \ref{fig7}.a give, for a human adult (70 kg) at rest, a value of 4 W for the total amount of heat extracted per unit of time from the lungs by the ventilation. It is approximately 5\% of the basal metabolic rate (BMR) of such an adult (around 80 W)\cite{Roza_1984}. At maximal effort and for $M = 70$ kg, a value of 62 W is calculated, which is approximately 4\% of the maximum metabolic rate (MMR) of a human adult (around 1400 W)\cite{Weibel2005}. As we have initially assumed that flow rate $Q$ is proportional to the BMR at rest and to the MMR at maximal effort, these observations can be qualitatively extended to the whole class of mammals (see the insert in Figure \ref{fig7}.a).

Regarding the hydric balance of the body, it is often assumed that the body of mammals is composed of approximately 60\% of water and that the loss of 1\% of this water leads to a slight state of dehydration\cite{Riebl}. Figure \ref{fig7}.a shows that the sole water extraction from the lungs by the ventilation has a limited contribution to dehydration. For instance, the volume of water extracted from the lungs of a human adult of 70 kg during 10 minutes at maximal effort is around 15 ml, remaining thus significantly smaller than the loss of 1\% of the water in the body of this human ($\simeq 400$ ml). Note that 10 minutes is an order of magnitude of the time before exhaustion at maximal effort for an accomplished sportsman \cite{Billat94}. 

Although it appears that the total amounts of water and heat extracted from the lungs by ventilation remain limited, the same conclusion is not reached if we look at the evaporation rate of water in the trachea, $J_1$. Indeed, the results presented in Figure \ref{fig7}.b show that, at maximal effort, this evaporation rate is between 10 and 20 $\mu$m min$^{-1}$. As the ASL has a thickness of approximately 10 $\mu$m, this evaporation would, in the absence of any balance mechanism, completely dehydrate the ASL in less than one minute. This is not anecdotal because, in humans, it has been shown that significant dehydration of the ASL leads, for example, to airway damage, impaired mucociliary clearance or to the release of inflammatory mediators \cite{Karamaoun22,Banner_1984,Kippelen_2013}. 

Regarding this balance mechanism, the secretion rate of the serous cells in the first generations of the lungs has been characterized in some research works. In a review dedicated to the regulation of the depth of the ASL and its composition\cite{Widdi}, a maximal replenishment rate of the ASL close to 20 $\mu$m min$^{-1}$ is mentioned, based on data obtained on bovines and cats. Later in the same review, the author mentions a study on tracheae of bovines\cite{Widdi2}, giving a replenishment rate between 10 and 35 $\mu$m min$^{-1}$. Thus, we can observe that these values of the maximal replenishment rate of the ASL by the serous cells are quite close to the calculated values of the evaporation rate in the trachea at maximal effort. It is an interesting result, showing that the presence of these cells in the first generations of the lungs of the mammals is an essential feature to allow, by keeping the mucus wet despite of the possible important evaporation of water, the proper operation of the mucociliary clearance. This also shows the importance of the significant condensation during expiration: at maximal effort, approximately 50\% of the water extracted during inspiration is returned to the mucosa during expiration (see Figure \ref{fig6}.b). Hence, without this process, the calculated values of $J_1$ would have been about twice larger than those presented in Figure \ref{fig7}.b. 

Finally, as detailed in Section 5 of the SM, an general analytical approximation for $J_1$ can be obtained by following the same assumptions as those made to obtain the analytical approximation of the effectivity (notably considering Pe$_i/\beta^2 \gg 1$):
\begin{equation}\label{analJgeneral}
J_1 \simeq J_1^* =  \sqrt{\frac{Q\mathcal{D}}{8 \pi \beta R_1^3}} \frac{(\eta^*)^2}{2-\eta^*}(c_{\text{sat}}(T_b) - c_{0}^{\text{insp}})
\end{equation}
with thus $\eta^*$ given by Equation \ref{analrend}. 

Using the allometric scaling laws given in Table \ref{allometriclaws}, $J_1^*$ can be expressed as a function of the mass $M$, at rest or at maximal effort. It appears that $J_1^* \propto M^{-1/8}$ at rest and $J_1^* \propto M^{-1/16}$ at maximal effort. As shown by Figure \ref{fig7}, $J_1^*$ appears very close to the values calculated by the model, except for the small mammals (with the analytical approximation unable to reproduce the non-monotonic behavior of $J_1$). This was expected as this analytical approximation is obtained for Pe$_i/\beta^2 \gg 1$.\\

\section*{Conclusion}

We have presented a comprehensive mathematical model of water and heat exchanges in the lungs of the terrestrial mammals. The results of the model shows that, on inspiration, the air is fully conditioned before leaving the bronchial region of the lungs (the small airways are "super-conditioners") and that, on expiration, significant condensation of water occurs in the lungs, whatever the mass of the mammal or the intensity of an effort. This condensation, markedly limiting the hydric loss (see Figures \ref{fig3} and \ref{fig6}), is a consequence of the low ability of the vascularization to supply with heat the tissues surrounding the bronchial epithelium, when compared to the intensity of the heat extraction to evaporate water in the ASL during inspiration. It is interesting to note that a similar mechanism is also described at the level of the turbinates of the nasal cavities of the mammals \cite{Sherwood_2012}.

The model also shows that two elements lead to significant differences in behavior between the small and the large mammals, regarding heat and water exchanges in their lungs. First, there is a change of regime of mass transfer in the lumen between the large and the small mammals, from the "boundary layer" regime for the large mammals to the "constant Sherwood" regime for the small ones. Second, the calculations show that the generation in which the largest amount of water or heat is extracted is the trachea as long as the mass is below a certain threshold (around 5 kg at rest, which corresponds to a cat, and around 50 g at maximal effort, which corresponds to a weasel). Then, when the mass is above this threshold, this generation gradually progresses into the bronchial tree with an increase of the mass. A scaling has been derived to evaluate how this generation progresses into the lungs when the mass or the intensity of an effort increases (Equation \ref{eqanalimax}). 

When the mass is above the above-mentioned threshold value and despite the complexity of the water and heat exchanges in the lungs, 
it appears that, independently of the mass, approximately 40\% of the water evaporated on inspiration is condensed on expiration at rest, while 50\% is at maximal effort. This independence on the mass is due to a subtle interplay between several phenomena: dependence of $\Lambda_i$ on the mass and on the generation index, progression into the lungs of the generation in which the largest amount of water is extracted. Consequently, except for the small mammals, the total amounts of water and heat extracted per unit of time from the lungs by the ventilation are almost proportional to $M^{3/4}$ at rest and to $M^{7/8}$ at maximal effort, such as the inspiration flow rate does. Analytical approximations have been derived to approximate these amounts (Equation \ref{totwextranal}), as well as the evaporation rate in the trachea (Equation \ref{analJgeneral}), the latter scaling with the mass as $M^{-1/8}$ at rest and as $M^{-1/16}$ at maximal effort. Please remind that the calculations show that the the total amounts of water and heat extracted per unit of time from the lungs by the ventilation remain limited (when compared to relevant quantities), which is not the case of the local evaporation rate that can become critical, especially at maximal effort and in the first generations of the lungs. 

Regarding the small mammals and for the two reasons mentioned above, the results clearly show a change of behavior: drop in the local and overall effectivities, deviations from the scaling laws with notably the presence of a maximum for the evaporation rate in the trachea as a function of the mass (see Figure \ref{fig7}).

Finally, it is striking to see that the results show that the lungs appear to be perfectly designed to fully condition the air at maximal effort (and clearly "over-designed" to achieve it at rest, except for the smallest mammals). Indeed, as shown in Figure \ref{fig3}.b, all the generations of the bronchial region of the lungs are mobilized for the conditioning at maximal effort (i.e. evaporation is taking place on the entire ASL--lumen interface). Moreover, as shown at the end of the results section, the calculated values of the evaporation rate of water from the bronchial mucosa at maximal effort can be very close to the maximal ability of the serous cells to replenish this mucosa with water; and please keep in mind the important role of condensation at expiration to limit this evaporation rate (by a factor 2, as shown by Figure \ref{fig6}.b). Even if these are preliminary results that deserve to be refined, they have at least the merit of raising a question: have some features of the lungs, especially the density of the serous cells (in the bronchial epithelium and the submucosal glands) and their ability to secrete mucus, as well as the vascularization of the bronchial wall (responsible for this condensation at expiration), been balanced by the evolution to ensure a proper conditioning at maximal effort?

\bibliography{biblio2}



\section*{Author contributions statement}
Conceptualization: all authors; methodology: all authors; model and code development: all authors; investigation: all authors; writing--original draft preparation: B.H.; writing--review and editing: all authors.

\section*{Data availability statement}
All data generated or analyzed during this study are included in this published article (and its Supplementary Information files).

\section*{Conflicts of Interest}
The authors declare that they have no known competing financial interests or personal relationships that could have appeared to influence the work reported in this paper.

\newpage

\section*{Nomenclature}

\begin{tabbing}

\textit{Roman symbols}\\

\=$c$\hspace{2cm}		\= Concentration of water vapor in air, $\text{mol} \, \text{m}^{-3}$\\
\>$d_{\text{alv}}$	 	\> Alveolar diameter, m\\
\>$\mathcal{D}$	 	\> Diffusion coefficient of water in air, $\text{m}^2 \, \text{s}^{-1}$\\
\>$E_h$				\> Amount of heat extracted per unit of time, $\text{W}$ \\
\>$E_w$				\> Amount of water extracted per unit of time, $\text{mol} \, \text{s}^{-1}$ \\
\>$i$					\> Generation index, - \\
\>$h$				\> Ratio of the radii of airways in two successive generations, - \\
\>$J$				\> Evaporation rate of water, $\text{mol} \, \text{m}^{-2} \, \text{s}^{-1}$ \\
\>$L$				\> Length of the airway, m \\
\>$\mathcal{L}$			\> Molar latent heat of vaporization, $\text{J} \, \text{mol}^{-1}$ \\
\>$M$					\> Mass of the body, kg \\
\>$n$				\> Number of generations in the bronchial region of the lungs, -\\
\>Pe					\> Peclet number, - \\
\>$Q$				\> Inspiration flow rate, $\text{m}^3 \, \text{s}^{-1}$ \\
\>$r$					\> Radial coordinate, - \\		
\>$R$				\> Radius of the airway, m \\			
\>Sh					\> Sherwood number of the mass transfer between the ASL and the lumen, - \\
\>$t_{\text{bl}}$			\> Characteristic time of blood renewal, s\\ 
\>$T$				\> Temperature, K\\ 
\>$U$				\> Average axial velocity of the air in the airway,  $\text{m} \, \text{s}^{-1}$ \\
\>$v$				\> Velocity of the air, -\\
\>$z$				\> Axial coordinate, - \\	[6pt]

\textit{Greek symbols}\\
	
\>$\alpha_w$			\> Thermal diffusivity of liquid water,  $\text{m}^{2} \, \text{s}^{-1}$\\	
\>$\beta$				\> $L/R$ ratio of the airway, -\\
\>$\eta$				\> Effectivity of extraction, -\\
\>$\lambda_w$			\> Thermal conductivity of liquid water,  $\text{W} \, \text{m}^{-1} \, \text{K}^{-1}$\\	
\>$\Lambda$			\> Dimensionless number, see Equation \ref{lambda}\\	
\>$\phi$ 				\> Factor accounting for the increase of the cardiac flow rate during a possible effort, -\\
\>$\Gamma$			\> Dimensionless number, see Equation \ref{gamma}\\	
\>$\Theta$			\> Dimensionless constant, see Equation \ref{analtheta} \\
\>$\psi$ 				\> Factor accounting for the increase of the inspiration flow rate during an effort, -\\ [6pt]

\textit{Subscripts and superscripts}\\
\>*					\> Analytical approximation\\
\>$\bar{c}$			\> Velocity-average of $c$\\
\>$\tilde{c}$			\> Dimensionless version of $c$\\
\>0					\> At the top of the trachea\\
\>$b$				\> Body\\
\>exp				\> Related to the expiration\\
\>$i$					\> In generation $i$\\
\>insp				\> Related to the inspiration\\
\>lim					\> $\Gamma$ or $\Lambda$ for Pe$/\beta^2 \gg 1$\\
\>max				\> Maximum of extraction\\
\>$r$					\> Component along the $r$ axis\\
\>ref					\> At the reference mass $M_{\text{ref}}$\\
\>sat					\> Saturated with water\\
\>sml				\> $\Gamma$ for small Pe$/\beta^2$\\
\>$z$				\> Component along the $z$ axis\\
\>$\mu$				\> At the ASL--lumen interface\\ [6pt]

\end{tabbing}

\newpage

\begin{figure}
    \centering
   \includegraphics[width=\textwidth]{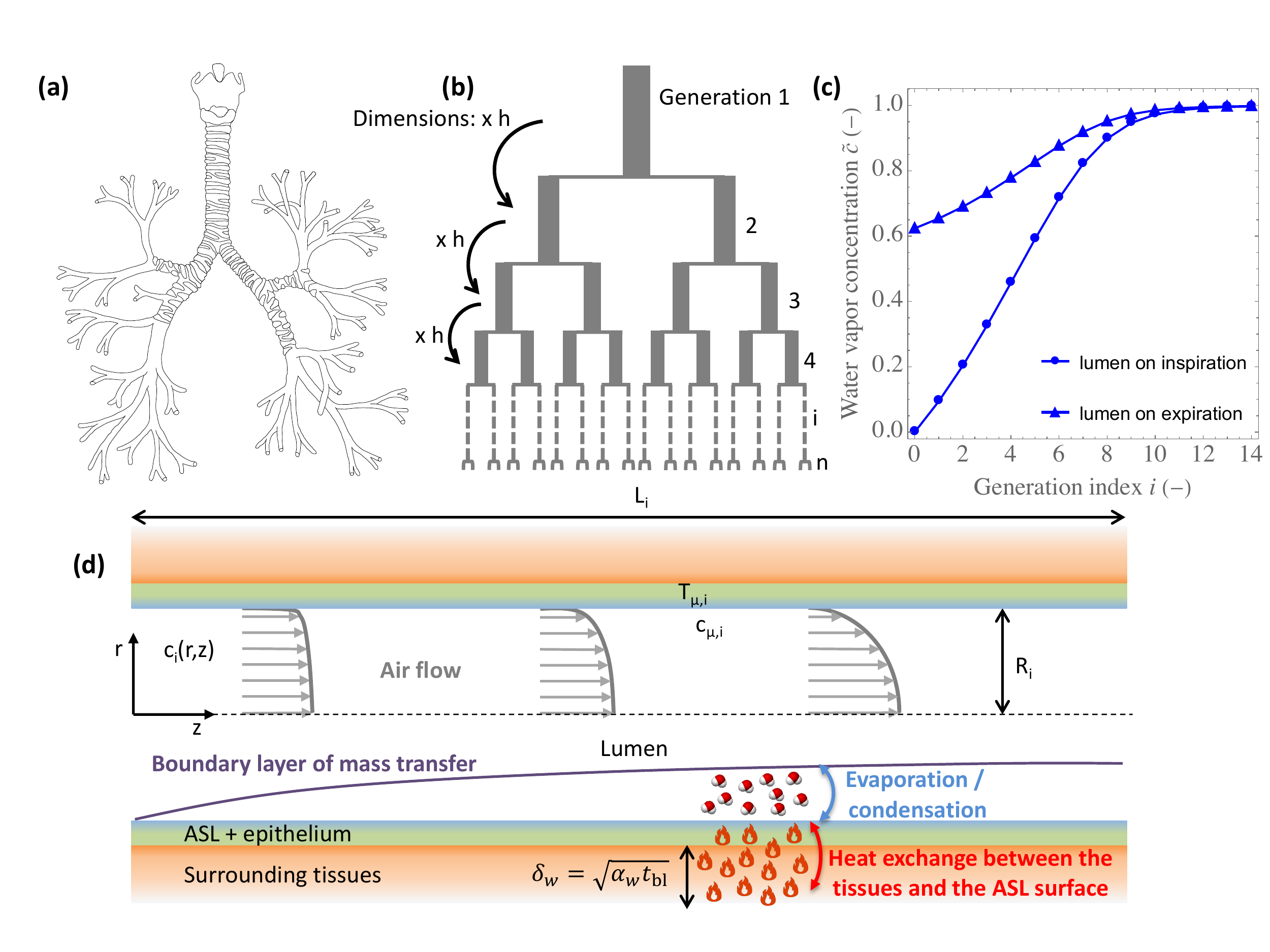}
   \caption{\textbf{(a)} Illustration of the bronchial region of the lungs, with the airways dividing successively into smaller ones. \textbf{(b)} "Weibel A" geometrical representation of the bronchial region of the lungs. \textbf{(c)} Example of water vapor concentration profiles, calculated with the model developed in this paper, within the lumen of the lungs of an adult human at rest. These concentrations are dimensionless: $\tilde{c} = 0$ is the concentration at the top of the trachea during inspiration and $\tilde{c} = 1$ is the saturation concentration of water in air at the body temperature. The circles represent the profile on inspiration. The triangles represent the expiration, and we see that significant condensation of water occurs during this expiration. 
\textbf{(d)} Sketch of an airway in the bronchial region of the lungs, showing the various phenomena involved in air conditioning: the heat exchange between the tissues and the surface of the ASL; the evaporation/condensation of water; the convective/diffusive transport of water vapor in the diffusion boundary layer developing on the ASL--lumen interface; and the establishment of the flow, from a "flat'' velocity profile to a parabolic one (if possible, according to the Reynolds number of the flow and the length of the airway).} 
 \label{general}
\end{figure}

\begin{figure}
    \centering
   \includegraphics[width=\textwidth]{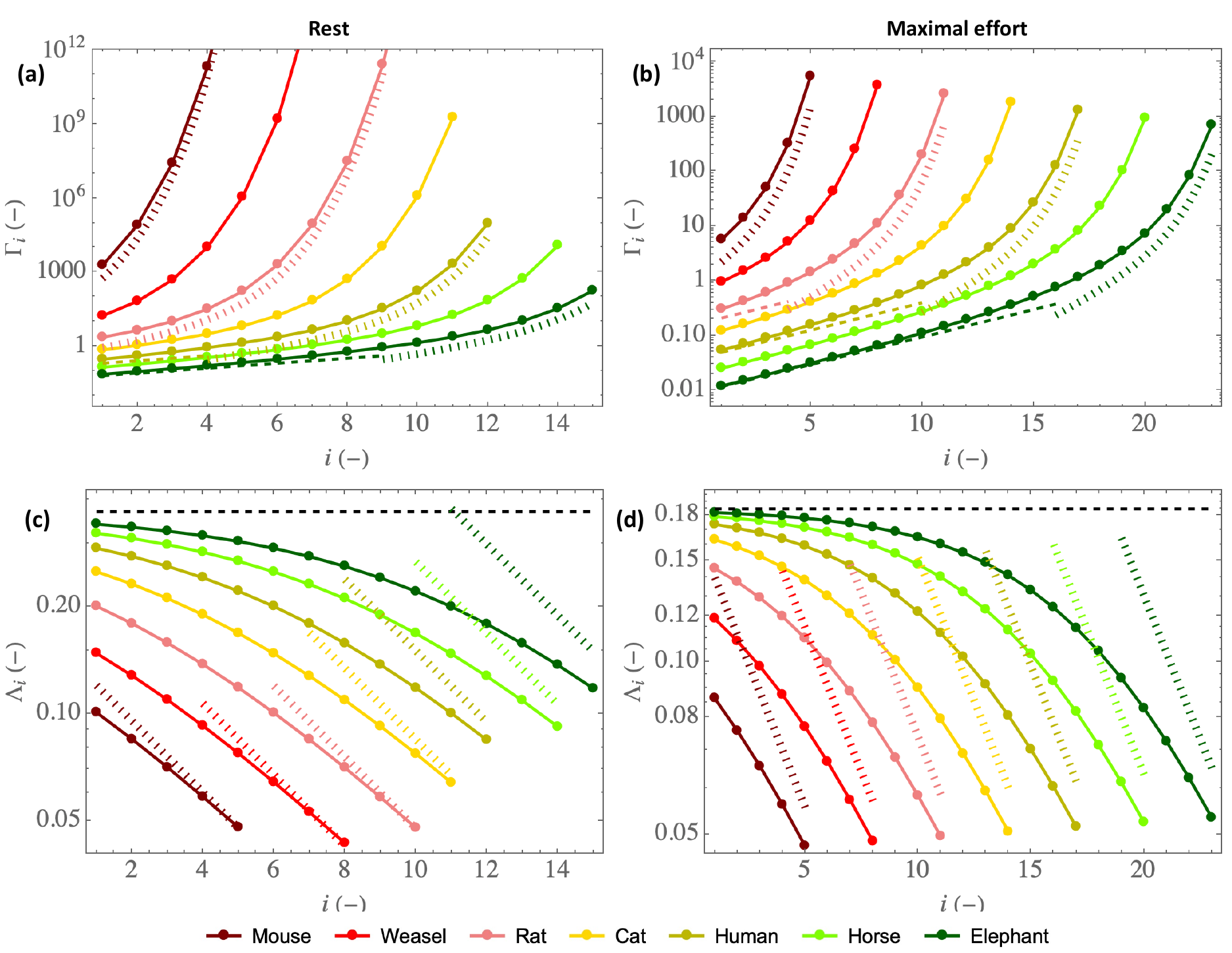}
   \caption{\textbf{(a-b)}: $\Gamma_i$ as a function of the generation index $i$, for mammals of different sizes, at rest \textbf{(a)} and at maximal effort \textbf{(b)}. The dashed and dotted curves provide values of $\Gamma_{i,\text{lim}}$ (Equation \ref{gammalim}) and $\Gamma_{i,\text{sml}}$ (Equation \ref{gammasmall}), respectively. \textbf{(c-d)}: $\Lambda_i$ as a function of the generation index $i$, for mammals of different sizes, at rest \textbf{(c)} and at maximal effort \textbf{(d)}. The dashed horizontal lines provide values of $\Lambda_{\text{lim}}$ (Equation \ref{lambdalim}), with $\phi/\psi = 1$ at rest and $\phi/\psi = 0.25$ at maximal effort (see Table \ref{allometriclaws}). The dotted lines provide values of $\Lambda_{i,\text{sml}}$ (Equation \ref{lambdasmall}).}
 \label{fig2}
\end{figure}

\begin{figure}
    \centering
   \includegraphics[width=\textwidth]{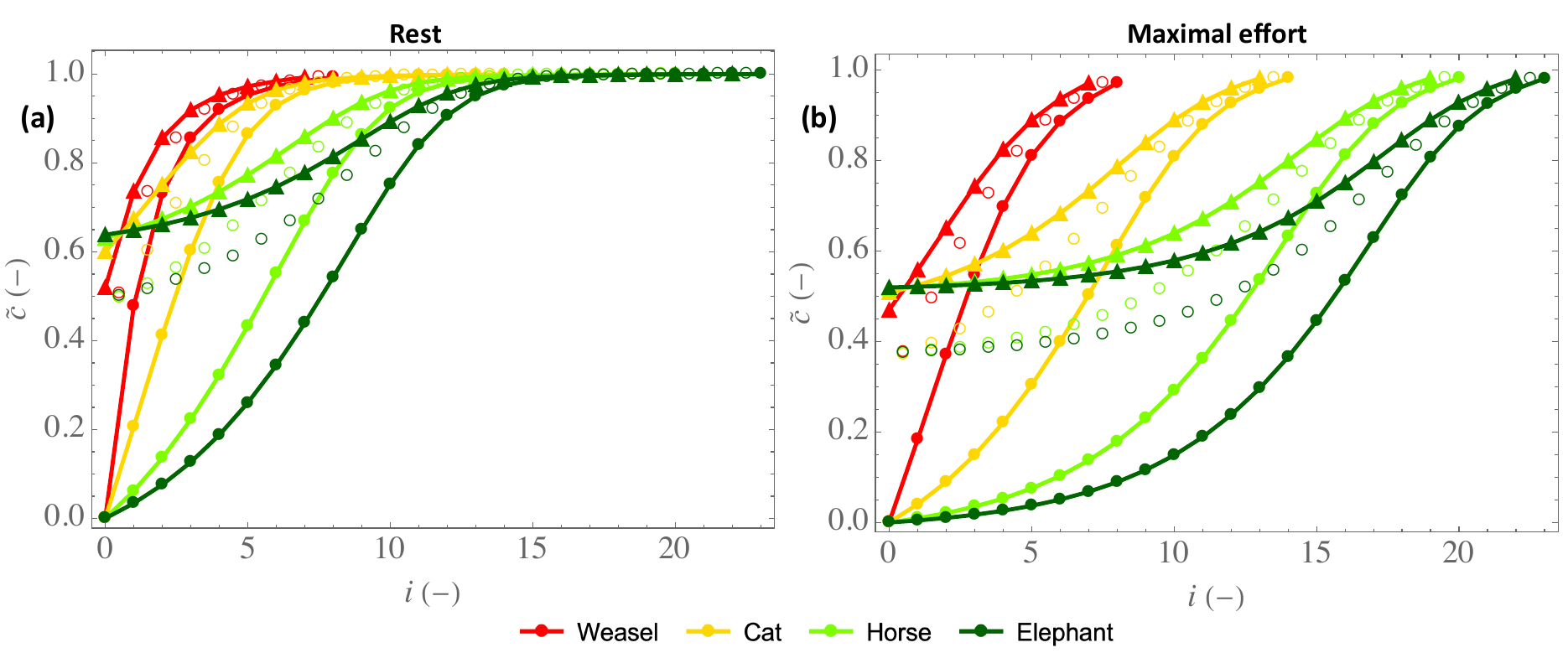}
   \caption{Dimensionless water concentration profiles in the lumen, during inspiration (full circles) and expiration (triangles), for mammals of different sizes, at rest \textbf{(a)} and at maximal effort \textbf{(b)}. Empty circles: values of $\tilde{c}_{\mu,i}$.}
 \label{fig3}
\end{figure}

\begin{figure}
    \centering
   \includegraphics[width=\textwidth]{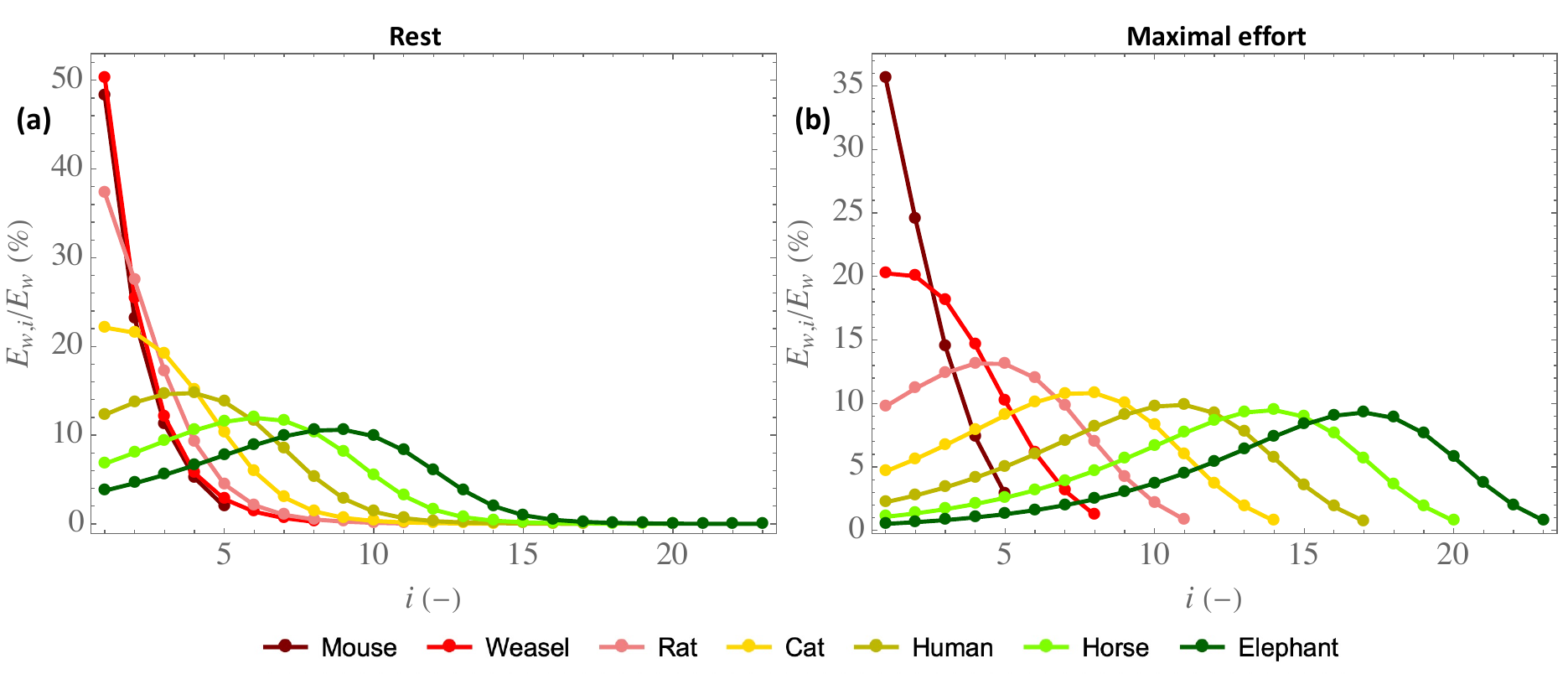}
   \caption{$E_{w,i}/E_w$ as a function of the generation index $i$, for mammals of different sizes, at rest \textbf{(a)} and at maximal effort \textbf{(b)}.}
 \label{fig4}
\end{figure}

\begin{figure}
    \centering
   \includegraphics[width=0.5\textwidth]{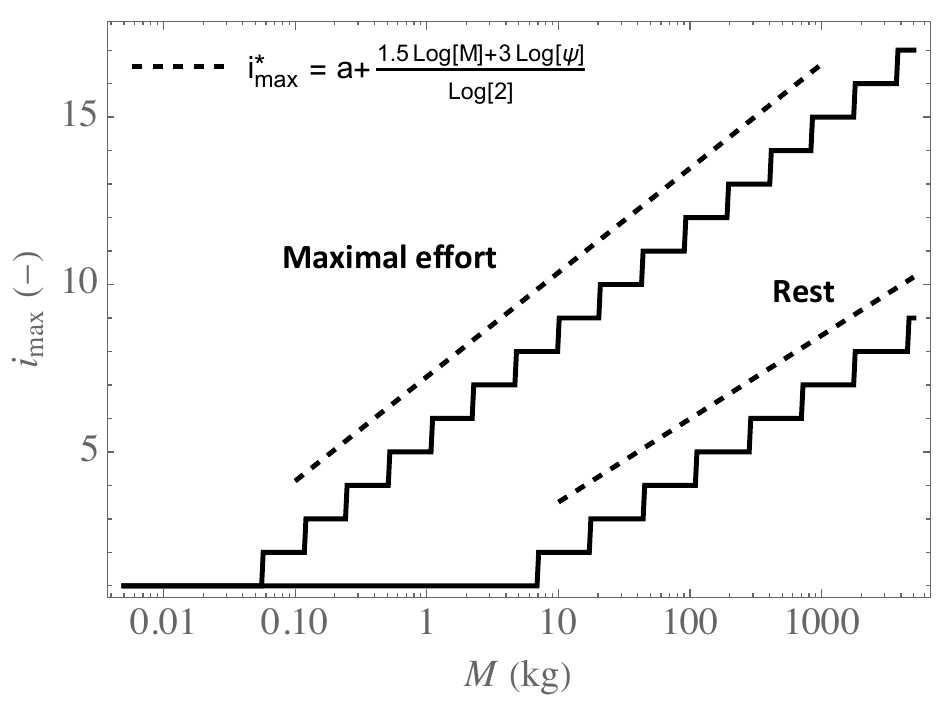}
   \caption{Index of the generation in which the largest amount of water is extracted $i_{\text{max}}$ as a function of $M$, at rest (upper continuous curve) and at maximal effort (lower continuous curve), as well as the scaling law given by Equation \ref{eqanalimax} (rest: lower dashed line, maximal effort: upper dashed line).}
 \label{fig5}
\end{figure}

\begin{figure}
    \centering
   \includegraphics[width=\textwidth]{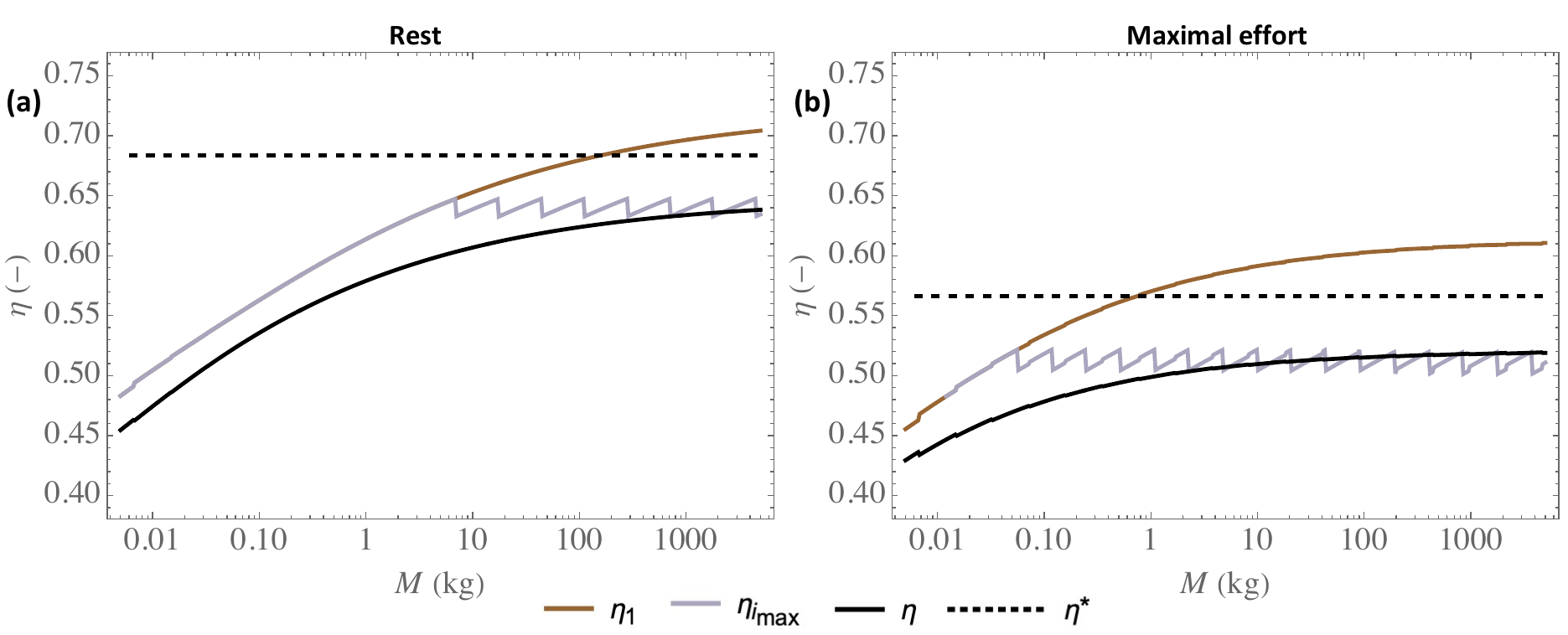}
   \caption{Effectivities of water (or heat) extraction as a function of the mammal mass $M$, at rest \textbf{(a)} and at maximal effort \textbf{(b)}. $\eta_1$: local effectivity in the trachea; $\eta_{i_{\text{max}}}$: $\eta_i$ for $i = i_{\text{max}}$; $\eta$: overall effectivity of water (or heat) extraction. $\eta^{*}$: analytical approximation of the effectivity (Equation \ref{analrend}).}
 \label{fig6}
\end{figure}

\begin{figure}
    \centering
   \includegraphics[width=\textwidth]{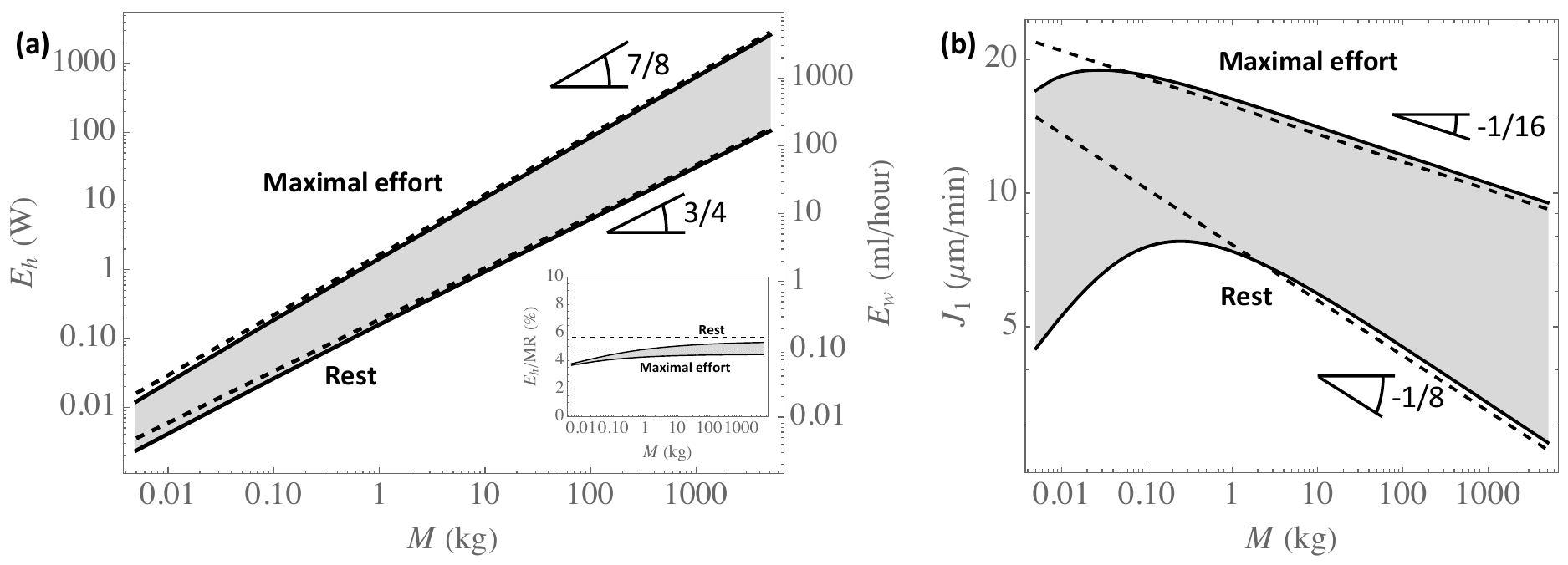}
   \caption{\textbf{(a)} Total amounts of water $E_w$ and heat $E_h$ extracted per unit of time from the lungs, as functions of $M$, at rest (lower continuous curve) and at maximal effort (upper continous curve). The two dashed lines give $E^*_w$ and $E^*_h$ (see Equation \ref{totwextranal}) at rest (lower line) and at maximal effort (upper line). The insert shows the ratio of $E_h$ at rest to the BMR (upper curve) and the ratio of $E_h$ at maximal effort to the MMR (lower curve). On this insert, the horizontal dashed lines give the ratio of $E^*_h$ at rest to the BMR (upper curve) and the ratio of $E^*_h$ at maximal effort to the MMR (lower curve). To compute these curves, we have assumed that a mammal of 70 kg has a BMR of 80 W and a MMR of 1400 W. \textbf{(b)} Time average of the evaporation rate in the trachea $J_1$ as a function of $M$, at rest (lower curve) and at maximal effort (upper curve). The two dashed lines give $J^*_1$ (see Equation \ref{analJgeneral}) at rest (lower line) and at maximal effort (upper line). $c_0^{\text{insp}} = 1.22$ mol m$^{-3}$ has been used to generate this figure.}
 \label{fig7}
\end{figure}

\newpage

\begin{table}\label{allometriclaws}
\small
\centering
\caption{\label{allometriclaws}Scaling with the mass $M$ of different parameters used in the model\cite{Kleiber,painter,Weibel2005,west_general_1997}, and their values at a reference mass $M_{\text{ref}} = 70$ kg. The reference values of $\beta$, $R_{1}$, $d_{\text{alv}}$ and $Q$ are those mentioned in several works\cite{Haut2021,karamaoun_new_2018,karamaoun_modeling_2016,Worthington91}. The reference value of $\psi$ can be calculated from correlations given by Weibel and Hoppeler \cite{Weibel2005} (for this purpose, the reference value of $\psi$ is assumed equal to the ratio of the maximum metabolic rate to the basal metabolic rate for a mammal of mass $M_{\text{ref}}$). The reference value of $\phi$ has been chosen based on data presented by Morris et al. \cite{Morris08}.}
\begin{tabular}{lccccc}
\textbf{Parameter} & \textbf{Symbol} &  \textbf{Scaling} &  \textbf{Value at $M = 70$ kg} &  \textbf{Unit}  \\ \hline
Alveolar diameter & $d_{\text{alv}}$ & $\propto M^{\frac{1}{12}}$ & 200 $\times 10^{-6}$ & m\\
Characteristic time of blood renewal in the tissues at rest & $t_{\text{bl}}$ & $\propto M^{\frac{1}{4}}$ & 2000 & s\\
Volumetric inspiration flow rate at rest & $Q$ & $\propto M^{\frac{3}{4}}$ & 250 $\times 10^{-6}$ & $\text{m}^3 \, \text{s}^{-1}$\\
Increase of the cardiac output at maximal effort  & $\phi$ & $\propto M^{\frac{1}{8}}$ &  4.5 & -\\
Increase of the inspiration flow rate at maximal effort  & $\psi$ & $\propto M^{\frac{1}{8}}$ & 18 & -\\
Radius of the trachea & $R_{1}$ & $\propto M^{\frac{3}{8}}$ & 7.5 $\times 10^{-3}$ & m\\
Ratio of the length to the radius of an airway  & $\beta$ & $\propto M^{-\frac{1}{8}}$  & 7 & -\\
\end{tabular}
\end{table}

\begin{table}\label{parametervalues}
\small
\centering
\caption{\label{parametervalues}Values of the parameters independent of the mass of the body. The different physicochemical properties are evaluated at the body temperature and, if relevant, for a relative humidity of 100\%, using data and equations given in the literature \cite{Lide2005,sobac_comprehensive_2015}.}
\begin{tabular}{lccc}
\textbf{Parameter} & \textbf{Symbol} &  \textbf{Value}  & \textbf{Units} \\ \hline
Derivative of the saturation concentration with respect to temperature & $\left . \frac{dc_{\text{sat}}}{dT}\right|_{T =T_b}$ & 0.12 & $\text{mol} \,\text{m}^{-3}  \,\text{K}^{-1}$  \\
Diffusion coefficient of water in air & $\mathcal{D}$ & 2.7 $\times 10^{-5}$ & $\text{m}^2 \, \text{s}^{-1}$  \\
Kinematic viscosity of the air & $\nu$ & 1.7 $\times 10^{-5}$ & $\text{m}^2 \, \text{s}^{-1}$\\
Molar latent heat of vaporization of water & $\mathcal{L}_m$ & 43470 & $\text{J} \, \text{mol}^{-1}$\\
Reference mass & $M_{\text{ref}}$ & 70 & kg \\
Saturation concentration of water in air, at $T_b = 310.15$ K & $c_{\text{sat}}(T_b)$ & 2.43 & $\text{mol} \,\text{m}^{-3}$\\ 
Size reduction at airways bifurcation & $h$ & $2^{-1/3}$ & - \\
Temperature of the body & $T_b$ & 310.15 & K \\
Thermal conductivity of the tissues surrounding the bronchial epithelium & $\lambda_w$ & 0.62 & $\text{W} \, \text{m}^{-1} \, \text{K}^{-1}$\\
Thermal diffusivity of the tissues surrounding the bronchial epithelium & $\alpha_w$ & 1.5 $\times 10^{-7}$ & $\text{m}^2 \, \text{s}^{-1}$\\
Volumetric mass of liquid water at the body temperature & $\rho_w$ & 993 & $\text{kg} \, \text{m}^{-3}$\\
\end{tabular}
\end{table}

\end{document}